\documentclass{emulateapj}
\usepackage{apjfonts}
\usepackage{graphicx}
\usepackage{amsmath}
\usepackage{amsfonts}

\shortauthors{}
\shorttitle{}

\begin{document}

% ------------------------------------------------------------------------
% New commands
%
\def\ltsima{$\; \buildrel < \over \sim \;$}
\def\lsim{\lower.5ex\hbox{\ltsima}}
\def\gtsima{$\; \buildrel > \over \sim \;$}
\def\gsim{\lower.5ex\hbox{\gtsima}}
\def\lam{\lambda=-1\fdg4 \pm 1\fdg1}

%\def\jk{JK}
% -------------------------------------------------------------------------
%

\title{
The Transit Light Curve project.~XIV.~Confirmation of Anomalous Radii\\
for the Exoplanets TrES-4\lowercase{b}, HAT-P-3\lowercase{b}, and WASP-12\lowercase{b}
}

\author{
Tucker Chan\altaffilmark{1},
Mikael Ingemyr\altaffilmark{1,2,3},
Joshua N.\ Winn\altaffilmark{1},
Matthew J.\ Holman\altaffilmark{4},\\
Roberto Sanchis-Ojeda\altaffilmark{1},
Gil Esquerdo\altaffilmark{4},
Mark Everett\altaffilmark{5}
}

\altaffiltext{1}{Department of Physics, and Kavli Institute for
  Astrophysics and Space Research, Massachusetts Institute of
  Technology, Cambridge, MA 02139, USA}

\altaffiltext{2}{Research Science Institute, Center for Excellence in Education, 8201 Greensboro Drive, Suite 215, McLean, VA 22102, USA}

\altaffiltext{3}{Present address:\ Silleg{\aa}rdsgatan 12, 686 95, V\"asrtra \"Amtervik, Sweden}

\altaffiltext{4}{Harvard-Smithsonian Center for Astrophysics, 60
 Garden Street, Cambridge, MA 02138, USA}

\altaffiltext{5}{Planetary Science Institute, 1700 E.~Fort Lowell Rd.,
  Suite 106, Tucson, AZ 85719}

\begin{abstract}

  We present transit photometry of three exoplanets, TrES-4b, HAT-P-3b, and WASP-12b, allowing for refined estimates of the systems' parameters. TrES-4b and WASP-12b were confirmed to be ``bloated'' planets, with radii of $1.706 \pm 0.056~\text{R}_{\rm Jup}$ and $1.736 \pm 0.092~\text{R}_{\rm Jup}$, respectively. These planets are too large to be explained with standard models of gas giant planets. In contrast, HAT-P-3b has a radius of $0.827\pm 0.055~\text{R}_{\rm Jup}$, smaller than a pure hydrogen-helium planet and indicative of a highly metal-enriched composition.  Analyses of the transit timings revealed no significant departures from strict periodicity. For TrES-4, our relatively recent observations allow for improvement in the orbital ephemerides, which is useful for planning future observations.

\end{abstract}

\section{Introduction}

A puzzling feature of the hot Jupiters is that many of them have radii that are either larger or smaller than one would have guessed prior to the discovery of this class of objects. This ``radius anomaly problem'' has been present since the first transiting planet was discovered by \citet{2000ApJ...529L..45C} and \citet{2000ApJ...529L..41H}, and still has no universally acknowledged resolution.  \citet{2010SSRv..152..423F} have reviewed many of the proposed solutions, and even in the short time since their review several other theories have been proposed \citep[see, e.g.,][]{2010ApJ...724..313P, 2010ApJ...714L.238B}. The small size of some planets can be explained as a consequence of heavy-metal enrichment, beyond the enrichment factors of Jupiter and Saturn and comparable to those of Uranus and Neptune. As for the larger radii, possible explanations include tidal friction, unexpected atmospheric properties, and resistive heating from electrical currents driven by star-planet interactions.

This paper presents follow-up observations of three exoplanets that were found to have anomalous radii. As in previous papers in this series, the purpose of the observations was to refine the system parameters (thereby checking on the magnitude of the radius problem) and to check for any transit timing anomalies that might be caused by additional gravitating bodies in the system.

Two of our targets are among the most ``bloated'' planets known. TrES-4 was discovered by \citet{2007ApJ...667L.195M}, and the two high-precision light curves that accompanied the discovery paper were reanalyzed by \citet{2008ApJ...677.1324T} and \citet{2009ApJ...691.1145S}. Here, we present 5 new light curves for this system. We also present 2 new light curves for WASP-12, a planet that was discovered by \citet{2009ApJ...693.1920H} and for which occultation photometry has been used to characterize the planet's atmosphere and orbit \citep{2010arXiv1003.2763C,2011Natur.469...64M}. Our third target, HAT-P-3 \citep{2007ApJ...666L.121T}, is in the opposite category of planets that are ``too small.'' \citet{2010MNRAS.401.1917G} have published 7 high-quality light curves of the system.  We present six new light curves, and provide independent estimates of the planetary and stellar parameters.

\section{Observations and Data Reduction}

Almost all the observations were conducted at the Fred Lawrence Whipple Observatory (FLWO) located on Mt.\ Hopkins, Arizona, using the 1.2m telescope and KeplerCam detector. The KeplerCam is a $4096^2$ CCD with a field of view of $23\farcm1 \times 23\farcm1$. The pixels were binned $2 \times 2$ on the chip for faster readout. The binned pixels subtend $0\farcs68$ on a side. Observations were made through Sloan $i$ and $z$ filters. One of the WASP-12 transits was observed with the Nordic Optical Telescope (NOT) located in the Canary Islands, using the ALFOSC detector.  The ALFOSC detector is a $2048^2$ CCD with a field of view of $6\farcm4 \times 6\farcm4$, corresponding to $0\farcs19$ per pixel. The observation was made through a Johnson $V$ filter. On each night we attempted to observe the entire transit, with at least an hour before ingress and an hour after egress, but the weather did not always cooperate.

\begin{figure*}[ht]
\begin{center}
\leavevmode
\hbox{
\epsfxsize=7.5in
\epsffile{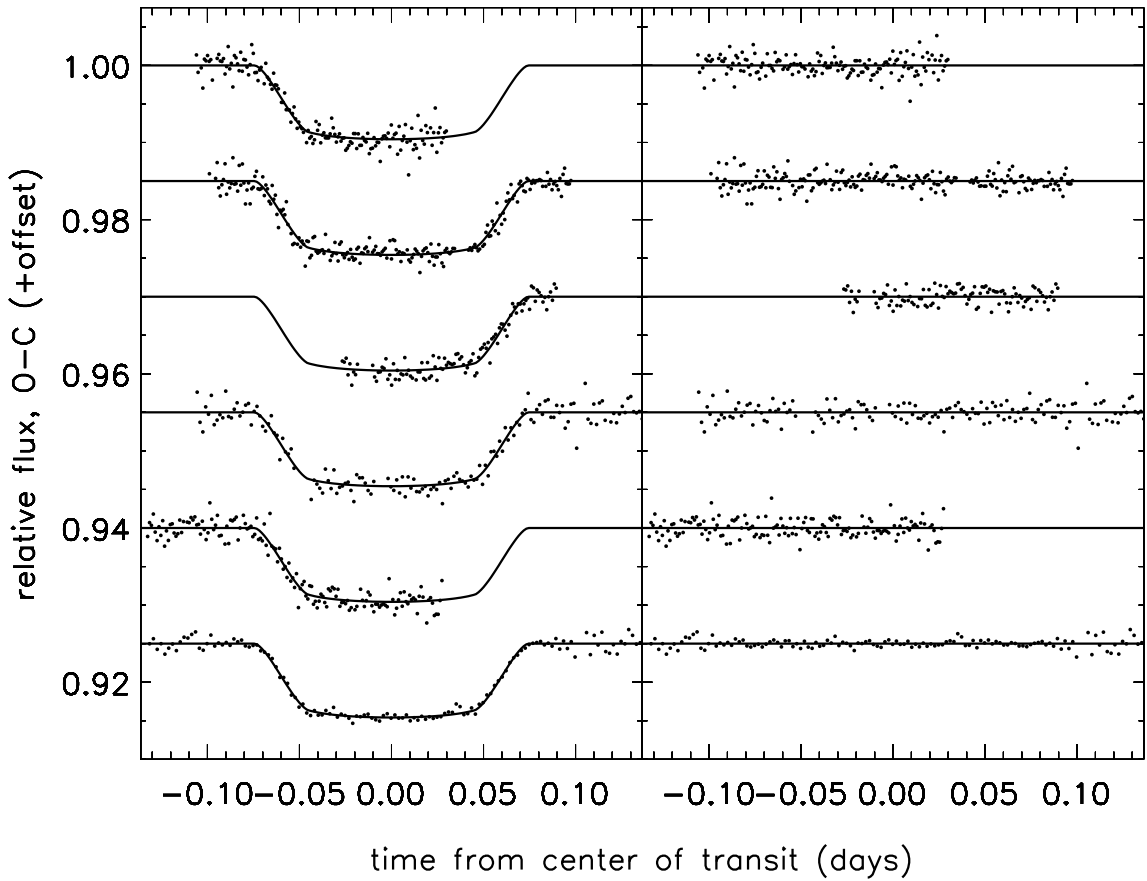}}
\end{center}
\vspace{-0.25in}
\caption{Relative photometry of TrES-4 in the $i$-band. From top to bottom,
the observing dates are
2008~Jun~10, 2009~Apr~1, 2009~May~3, 2010~Apr~27 and 2010~May~4.
See Table 1 for the cadence and rms residual of each light curve.
The bottom plot is a composite light curve averaged into 3~min bins.}
\vspace{0.0in}
\label{fig:lc-tres4}
\end{figure*}

We performed overscan correction, trimming, bias subtraction and flat-field division with IRAF\footnote{IRAF is distributed by the National Optical Astronomy Observatory, which is operated by the Association of Universities for Research in Astronomy (AURA) under cooperative agreement with the National Science Foundation.}. To generate the light curves, we performed aperture photometry on the target star and all the comparison stars with similar brightnesses to the target star (within about a factor of two).  We tried many different choices for the photometric aperture and found, unsurprisingly, that the best aperture diameter was approximately twice the full width at half maximum (FWHM) of the star.  A comparison signal was formed from the weighted average of the flux histories of the comparison stars. The weights were chosen to minimize the out-of-transit (OOT) noise level. Some comparison stars that did not seem to provide a good correction were rejected. In general, 6-8 comparison stars were used to generate the final comparison signal. The target star's flux history was divided by the comparison signal and then multiplied by a constant to give a mean flux of unity outside of the transit.

Table 1 is a journal of all our observations, including those that were spoiled by bad weather. The light curves are displayed in Figures 1--4, after having been corrected for differential extinction.  Section 3 explains how this correction was applied. The airmass-corrected data are given in electronic form in Tables 2-4.

\begin{figure*}[ht]
\begin{center}
\leavevmode
\hbox{
\epsfxsize=7.5in
\epsffile{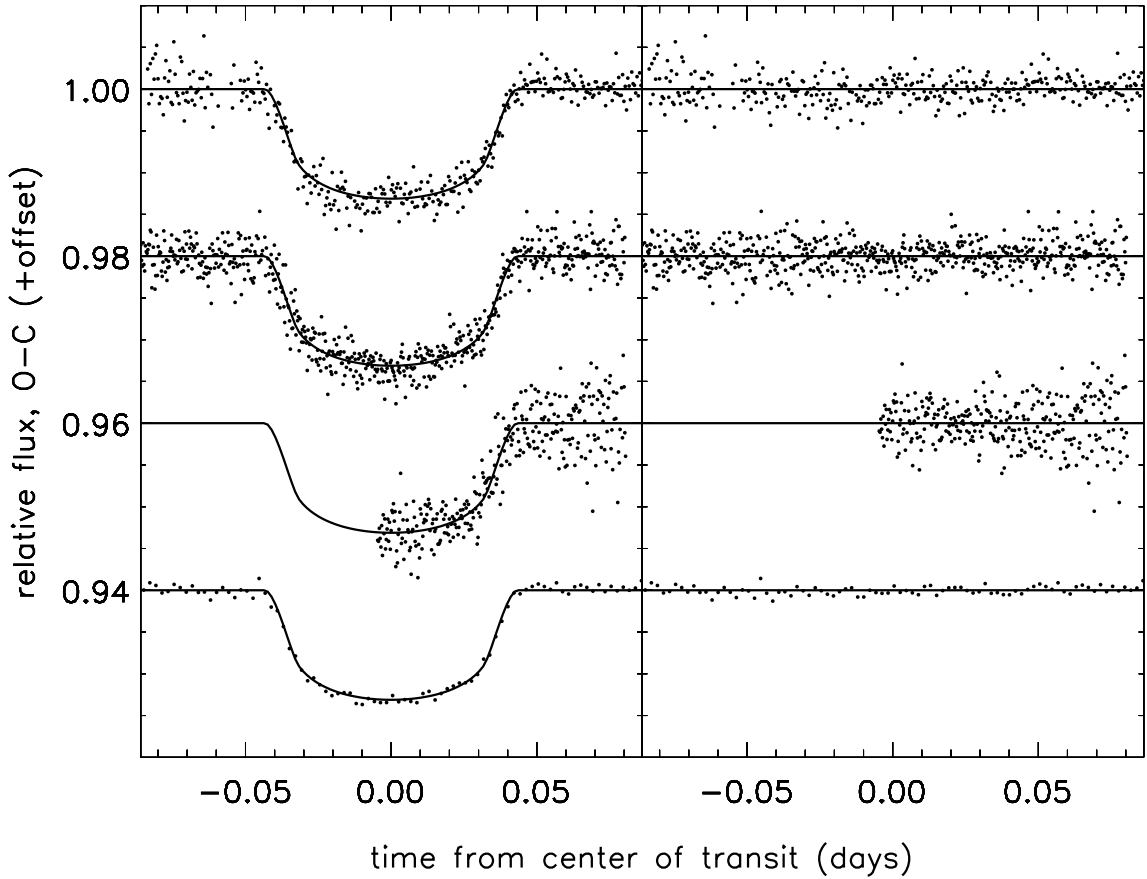}}
\end{center}
\vspace{-0.25in}
\caption{Relative photometry of HAT-P-3 in the $i$-band.
From top to bottom,
the observing dates are
2008~Mar~8, 2008~Apr~6 and 2008~May~5.
See Table 1 for the cadence and rms residual of each light curve.
The bottom plot is a composite light curve averaged into 3~min bins.}
\label{fig:lc-hat3-i}
\end{figure*}

\begin{figure*}[ht]
\begin{center}
\leavevmode
\hbox{
\epsfxsize=7.5in
\epsffile{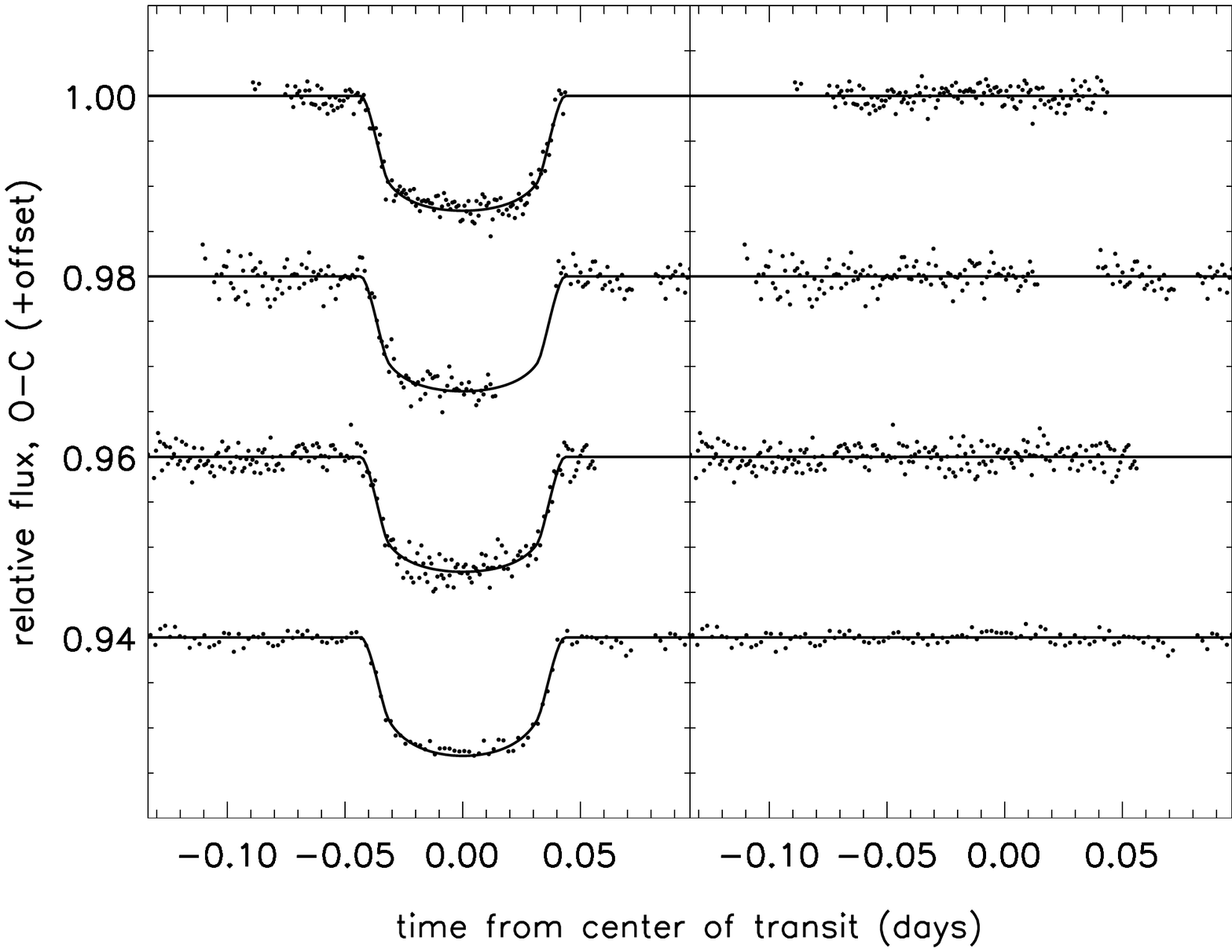}}
\end{center}
\vspace{-0.25in}
\caption{Relative photometry of HAT-P-3 in the $z$-band.
From top to bottom,
the observing dates are
2009~Mar~14, 2009~Mar~20 and 2009~Apr~15.
See Table 1 for the cadence and rms residual of each light curve.
The bottom plot is a composite light curve averaged into 3~min bins.}
\label{fig:lc-hat3-z}
\end{figure*}

\begin{figure*}[ht]
\begin{center}
\leavevmode
\hbox{
\epsfxsize=7.5in
\epsffile{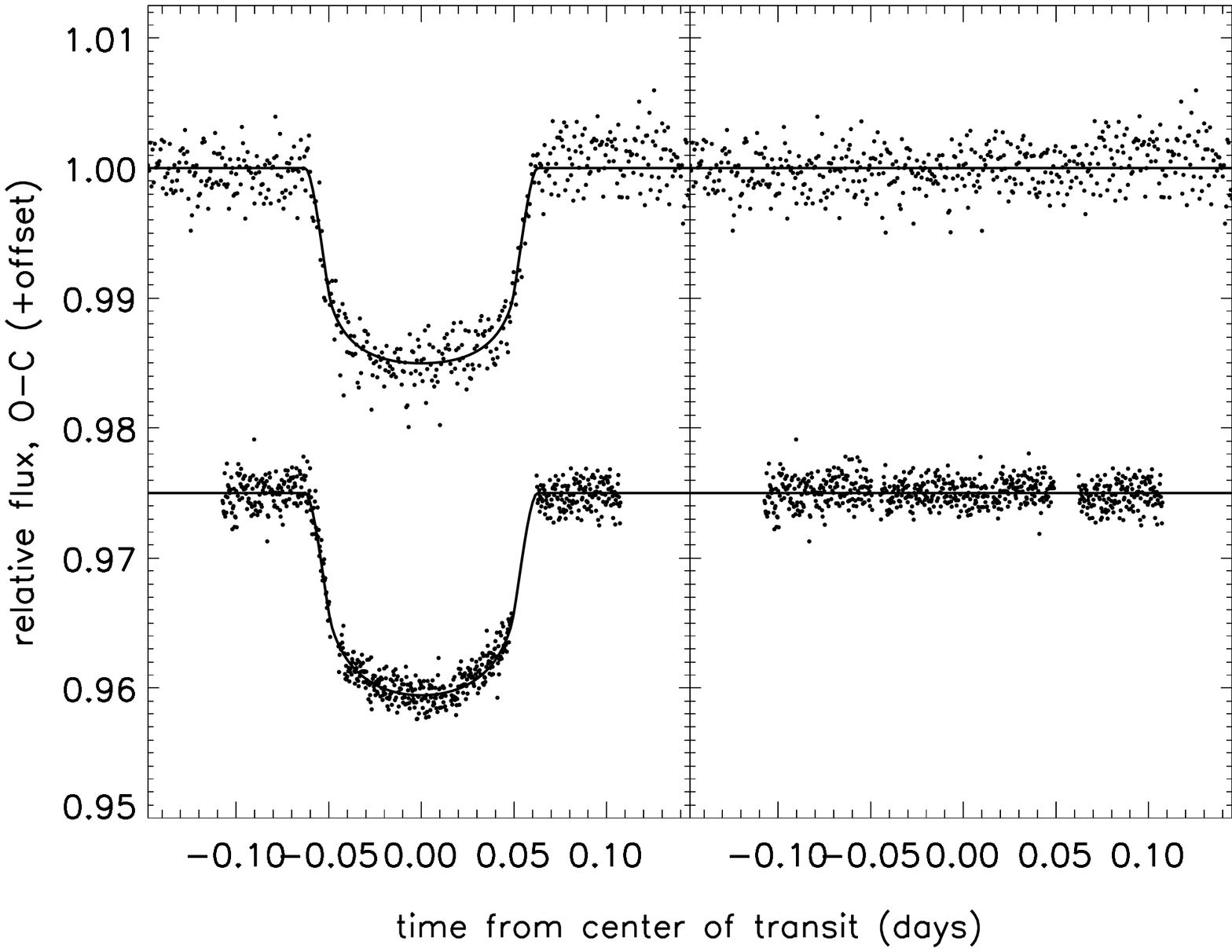}}
\end{center}
\vspace{-0.25in}
\caption{Relative photometry of WASP-12 in the $z$- and $V$-bands.
The top light curve is based on $z$-band observations on 2009~Jan~08.
See Table 1 for the cadence and rms residual of each light curve.
The bottom light curve is based on
$V$-band observations on 2009~Dec~06.}
\label{fig:lc-wasp12}
\end{figure*}

\section{Determination of System Parameters}
\label{sec:sys-params}

Our techniques for light-curve modeling and parameter estimation are similar to those
employed in previous papers in this series \citep[see, e.g.,][]{2006ApJ...652.1715H,2007AJ....133...11W}. The basis for the light-curve model was the formula of \citet{2002ApJ...580L.171M}, assuming
quadratic limb darkening and a circular orbit. The set of model parameters included the planet-to-star radius ratio ($R_p/R_\star$), the stellar radius in units of orbital distance ($R_\star/a$), the impact parameter ($b\equiv a\cos i/R_\star$, where $i$ is the orbital inclination), the time of conjunction for each individual transit ($T_c$), and the limb-darkening parameters $u_1$ and $u_2$.
We also fitted for two parameters ($\Delta m_0$, $k_z$) specifying a correction for differential extinction,
\begin{equation}
\label{eq:diff-ext}
\Delta m_{\rm cor} = \Delta m_{\rm obs} + \Delta m_0 + k_z z,
\end{equation}
where $z$ is the airmass, $\Delta m_{\rm obs}$ is the observed magnitude, and $\Delta m_{\rm cor}$ is the corrected magnitude that to be compared to the idealized transit model.

Since the data are not precise enough to determine both of the limb-darkening parameters, we followed the suggestion of \citet{2008MNRAS.390..281P} to form uncorrelated linear combinations of those parameters. We allowed the well-constrained combination to be a free parameter and held the poorly-constrained combination fixed at a tabulated value for a star of the appropriate type. In P{\'a}l's notation, the rotation angle $\phi$ was taken to be near $37^\circ$ in all cases. To determine the tabulated values we used a program kindly provided by J.\ Southworth to query and interpolate the tables of \citet{2004yCat..34281001C}.\footnote{The interpolated values $(u_1,u_2)$ for the cases of TrES-4 $i$, HAT-P-3 $i$, HAT-P-3 $z$, WASP-12 $z$, and WASP-12 $V$ are $(0.20,0.37)$, $(0.38,0.27)$, $(0.30,0.29)$, $(0.14,0.36)$, $(0.36,0.35)$, respectively.}

For parameter estimation, we used a Markov Chain Monte Carlo (MCMC) algorithm to sample from the posterior probability distribution, employing the
Metropolis-Hastings jump criterion and Gibbs sampling. Uniform priors were adopted for all parameters,
and the likelihood was taken to be $\exp(-\chi^2/2)$ with
\begin{equation}
\label{eqn:chisq}
\chi^2 = \sum_{i,j} \left( \frac{f_{{\rm obs},i,j} - f_{{\rm calc}, i,j}}{\sigma_{i,j}} \right)^2
\end{equation}
where $f_{{\rm obs}, i,j}$ is the $j$th data point from the $i$th light curve, $f_{{\rm calc},i,j}$ is the calculated light curve based on the current parameters, and $\sigma_{i,j}$ is the uncertainty associated with $f_{{\rm obs}, i,j}$.  All the light curves for a given planet were fitted simultaneously. The uncertainties were determined in a two-step process.  First, the standard deviation $\sigma_i$ of the OOT data was determined for each light curve.  In a few cases, the pre-transit and post-transit noise levels were very different (due to different airmasses); in these cases the starting point was a function $\sigma_{i,j}$ that interpolated linearly between the two differing noise levels. Second, the preceding uncertainty estimates were multiplied by a correction factor $\beta\geq 1$ intended to account for time-correlated noise. The $\beta$ factor was determined with the ``time-averaging'' procedure
\citep{2006MNRAS.373..231P,2008ApJ...683.1076W,2009ApJ...704...51C}, using bin sizes bracketing the ingress/egress duration by a factor of 2. The values of $\beta$ are given in Table~\ref{tbl:obs}.

The starting point for each Markov chain was determined by minimizing $\chi^2$, and then perturbing those parameters by Gaussian random numbers with a standard deviation of $10\sigma$, where $\sigma$ is the rough uncertainty estimate returned by the least-squares fit. We ran several test chains to establish the appropriate jump sizes, giving acceptance rates near 40\%. Then we ran $4-5$ chains each with $10^6$ links, ignored the initial 20\% of each chain, and ensured convergence according to the Gelman-Rubin statistic \citep{GR1992}.  The quoted value for each parameter is the median of the one-dimensional marginalized posterior, and the quoted uncertainty interval encloses 68.3\% of the probability (ranging from the 15.85\% to the 84.15\% levels of the cumulative probability distribution).

\section{Results}

The results for the model parameters, and various derived parameters of interest, are given in Tables~\ref{tbl:params-tres4},~\ref{tbl:params-hat3}~and~\ref{tbl:params-wasp12}. The next subsection explains how the stellar and plantetary dimensions were calculated from the combination of light-curve parameters and stellar-evolutionary models. This is followed by a subsection presenting an examination of the transit times.

\subsection{The stellar and planetary radii}

The stellar and planetary radii and masses cannot be determined from transit parameters alone. The route we followed to determining these dimensions was to set the mass scale by using an estimated stellar mass $M_\star$ and the observed semiamplitude $K_\star$ of the star's radial-velocity orbit. The stellar mass is itself estimated by using stellar-evolutionary models with inputs from the observed spectral parameters, as well as the mean density that is calculated from the light-curve parameters. For the relevant formulas and discussion see \citet{2007ApJ...664.1190S} or \citet{2010arXiv1001.2010W}. We used previously measured values of $K_\star$, documented in Tables~\ref{tbl:params-tres4},~\ref{tbl:params-hat3}~and~\ref{tbl:params-wasp12}.  For the evolutionary models, we used the Yonsei-Yale ($\rm Y^2$) isochrones \citep{2001ApJS..136..417Y}.

The $\rm Y^2$ isochrones can be thought of as an algorithm that takes as input the age, metallicity, mass, and concentration of $\alpha$-elements of a star, and returns the star's temperature, mass, density, and other properties. We interpolated the $\rm Y^2$ isochrones in age from $0.1$ to $14$ Gyr in steps of $0.1$ Gyr, and in metallicity from $-0.20$ to $0.58$ dex in steps of $0.02$ dex. Then we used linear interpolation to create a $4\times$ finer mass sampling for each metallicity and age. We assumed the concentration of $\alpha$-elements to be solar.
To each model star in the resulting Y$^2$ grid, we assigned a likelihood
based on the measured metallicity $Z$ and effective temperature $T_{\rm eff}$ (taken
from the literature) as well as the
stellar mean density $\rho_\star$ determined solely from the transit parameters.
Following \citet{2009ApJ...696..241C}, the likelihood was taken to be proportional
to $n \exp( - \chi^2/2)$,
where
\begin{equation}
\chi^2 = \left( \frac{Z-Z_{\rm obs}}{\sigma_Z} \right)^2 +
\left( \frac{T_{\rm eff} - T_{\rm eff, obs}}{\sigma_{T_{\rm eff}}} \right)^2 +
\left( \frac{\rho_\star}{\rho_{\star, {\rm obs}}} \right)^2,
\end{equation}
and $n$ is the number density of stars as a function of mass, according to Salpeter's law with exponent $-1.35$. The effect of multiplying by $n$ is to set a prior so that extremely rare and short-lived stars are disfavored, even though they might provide a good fit.  (The effect of this prior was generally small.) Finally, the ``best-fitting'' values and uncertainties were computed from the appropriate likelihood-weighted integrals in the space of model stars.

With the stellar mass thereby determined, it is possible to compute the other dimensions $R_p, R_\star,$ and $M_p$.  The results are given in Tables~\ref{tbl:params-tres4},~\ref{tbl:params-hat3}~and~\ref{tbl:params-wasp12}. We note that this procedure does not take into account any uncertainty in the $\rm Y^2$ isochrones themselves and, therefore, is subject to systematic errors that probably amount to a few percent \citep[see, e.g.,][]{2008ApJ...677.1324T}.

\subsection{Transit times and revised ephemerides}
\label{subsec:ephemeris}

We analyzed the new transit times in conjunction with previously published midtransit times, to seek evidence for significant discrepancies from strict periodicity and refine the ephemerides to allow for accurate prediction of future events. The new transit times are given in Tables~\ref{tbl:midtransits-tres4},~\ref{tbl:midtransits-hat3}~and~\ref{tbl:midtransits-wasp12}, with uncertainties determined by the MCMC analysis described in Section 3.  Previously published midtransit times were taken from \citet{2010MNRAS.401.1917G}, \citet{2009ApJ...693.1920H}, \citet{2009ApJ...691.1145S}, \citet{2007ApJ...666L.121T}, and \citet{2007ApJ...667L.195M}.

The transit times for each system were fitted with a linear function,
\begin{equation}
\label{eqn:linfit}
T_C = T_0 + E \cdot P
\end{equation}
where $E$ is an integer (the epoch), $P$ is the period, and $T_0$ is a particular reference time. The best-fitting values of $P$ and $T_0$ were determined by linear regression. Figures 5-7 show the residuals to the fits, and the captions specify the minimum $\chi^2$ and the number of degrees of freedom. In no case is there clear evidence of timing anomalies. For TrES-4, there is formally only a 6\% chance of obtaining such a large $\chi^2$ with random Gaussian errors, but we do not deem this significant enough to warrant special attention.

The case of WASP-12 required somewhat special treatment because \citet{2009ApJ...693.1920H} did not report individual midtransit times, but rather a consensus reference time based on observations of multiple transits. We included their quoted reference time as a single data point.  \citet{2010arXiv1003.2763C} provided many transit times obtained over several years, but most of the data were from amateur observers and were not presented in detail or evaluated critically. For this reasons we did not include them; however, as a consequence, there are not many points in our fit.

The results for $P$ and $T_0$ are given in Tables 5-7. They are based on a fit in which all of the uncertainties of the transit times were rescaled by a common factor to give $\chi^2/N_{\rm dof} = 1$. Our intention is to provide conservative error estimates to allow for planning of future observations. The uncertainties on the individual transit times given in Tables 8-10 were {\it not} rescaled in this way, nor were the error bars that are plotted in Figures 5-7.

\begin{figure}[ht]
\epsscale{0.8}
\plotone{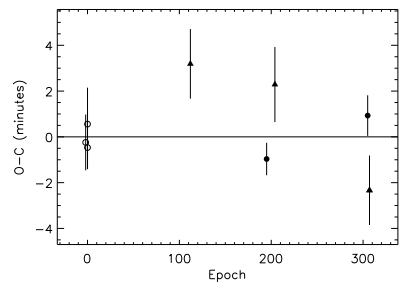}
\caption{  Timing residuals for TrES-4. The data presented
in this paper are labeled with solid circles (complete transits) and solid
trianges (partial transits).
Open circles represent transits observed by \citet{2007ApJ...667L.195M}.
The best fit gives $\chi^2=12.1$ with 6 degrees of freedom.
The probability of obtaining a higher $\chi^2$ with random Gaussian
data points is about 6\%.
\label{fig:timing-residuals-tres4}}
\end{figure}

\begin{figure}[ht]
\epsscale{0.8}
\plotone{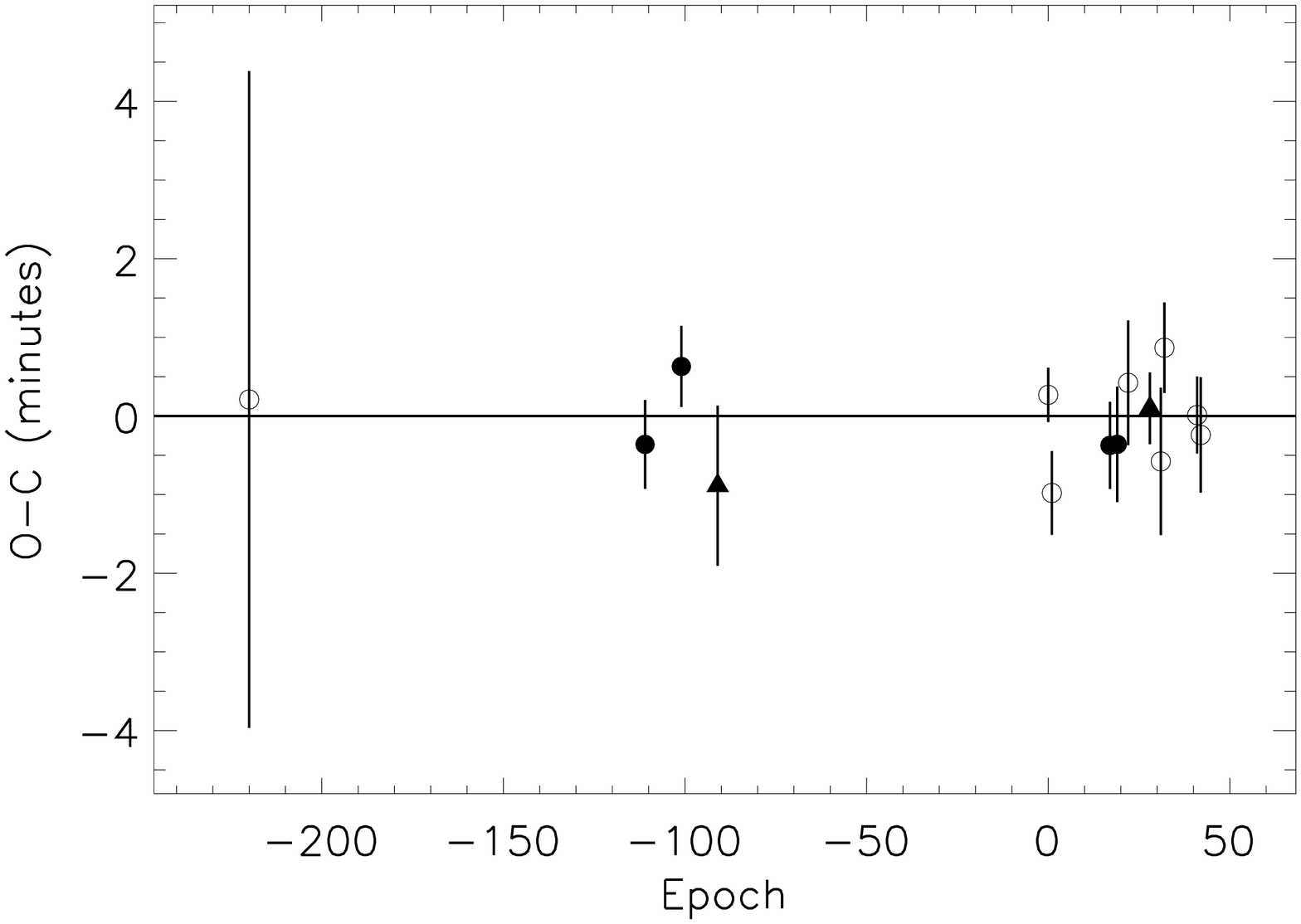}
\caption{ Timing residuals for HAT-P-3.
The data presented
in this paper are labeled with solid circles (complete transits) and solid
trianges (partial transits).
Open circles represent transits observed by \citet{2010MNRAS.401.1917G, 2007ApJ...666L.121T}.
The best fit gives $\chi^2=10.4$ with 12 degrees of freedom.
The probability of obtaining a higher $\chi^2$ with random Gaussian
data points is about 60\%.
\label{fig:timing-residuals-hat3}}
\end{figure}

\begin{figure}[ht]
\epsscale{0.8}
\plotone{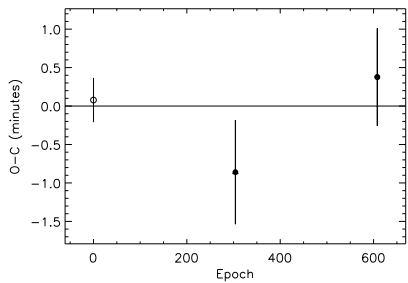}
\caption{ Timing residuals for WASP-12.
Solid circles are complete transits that we observed, and the open circle
is the reference transit time (derived from observations of multiple events) quoted
by \citet{2009ApJ...693.1920H}. The best fit gives $\chi^2=2.0$ with 1 degree of freedom.
\label{fig:timing-residuals-wasp12}}
\end{figure}

% \vspace{0.1in}

\section{Summary and Discussion}

We have presented new photometry and new analyses of the transiting exoplanets TrES-4b, HAT-P-3b, and WASP-12b.  Whereas the discovery papers reporting TrES-4b and HAT-P-3b included only a few high-precision light curves, our analyses are based on 5-6 such datasets. Likewise, the WASP-12b discovery paper featured only one high-precision light curve, to which we have added two. We have applied consistent and conservative procedures for parameter estimation, including an accounting for uncertainties in the limb-darkening law and due to time-correlated noise, as well as linkage between the light curve parameters and stellar-evolutionary models, that were not always applied by previous authors. In the past these efforts have occasionally led to significant revisions of the planetary dimensions \citep[see, e.g.][]{2007AJ....134.1707W,2008ApJ...683.1076W}.

In the present case our results are in agreement with the previously reported results. All of the TrES-4 parameters agree to within 2$\sigma$ with the results reported by \citet{2008ApJ...677.1324T}. Two of the most important parameters, the mass and radius of the planet, agree to within 1$\sigma$. Our HAT-P-3 parameters agree to within 2$\sigma$ with those reported by \citet{2010MNRAS.401.1917G}, and the planetary mass and radius agree to within 1.3$\sigma$.  Our WASP-12 parameters agree to within 2$\sigma$ with those reported by \citet{2009ApJ...693.1920H}, and the mass and radius of the planet agree to within 1$\sigma$.

Our transit ephemeris for TrES-4 agrees with that of \cite{2007ApJ...667L.195M}, and the refined orbital period is about 20 times more precise.  Our ephemeris for HAT-P-3 agrees with that of \citet{2010MNRAS.401.1917G}, and is 2-3 times more precise.  Our ephemeris for WASP-12 agrees with that of \citet{2009ApJ...693.1920H}, and is of comparable precision, despite being based on only a few well-documented data points.

The reason that TrES-4, HAT-P-3, and WASP-12 are of particular interest is because their measured dimensions do not agree with standard models of gas giant planets.  TrES-4 and WASP-12 are heavily bloated, with radii too large for their masses, while HAT-P-3 is too small. With reference to the tables of \citet{2007ApJ...659.1661F}, TrES-4 and WASP-12 are incompatible with pure hydrogen-helium giant planets at the 10$\sigma$ and 5$\sigma$ levels, respectively.  Enhancing these planets with metals to the degree of Jupiter or Saturn would only make the problem worse.  HAT-P-3, in contrast, is compatible with the tabulated models if it is endowed with approximately 100~$M_\earth$ of heavy elements. Our results do not change these interpretations of the three systems we have studied. Rather, our results lend more confidence to the claims that the dimensions of the planets are anomalous, and merit attention by theoreticians who seek to solve the radius anomaly problem.

\acknowledgments We thank Gerald Nordley for checking some of the entries in Tables 5-7. We gratefully acknowledge support from the NASA Origins program through award NNX09AB33G and from the Research Science Institute, a program of the Center for Excellence in Education. Some of the data presented herein were obtained with the Nordic Optical Telescope (NOT), operated on the island of La Palma jointly by Denmark, Finland, Iceland, Norway, and Sweden, in the Spanish Observatorio del Roque de los Muchachos Instituto Astrofisica de Canarias, and ALFOSC, which is owned by the Instituto Astrofisica de Andalucia (IAA) and operated at the NOT under agreement between IAA and NBlfAFG of the Astronomical Observatory of Copenhagen.

\bibliographystyle{apj}
\bibliography{apj-jour,myrefs}

\begin{thebibliography}{29}
\expandafter\ifx\csname natexlab\endcsname\relax\def\natexlab#1{#1}\fi

\bibitem[{{Batygin} \& {Stevenson}(2010)}]{2010ApJ...714L.238B}
{Batygin}, K., \& {Stevenson}, D.~J. 2010, \apjl, 714, L238

\bibitem[{{Campo} {et~al.}(2010){Campo}, {Harrington}, {Hardy}, {Stevenson},
  {Nymeyer}, {Ragozzine}, {Lust}, {Anderson}, {Collier-Cameron}, {Blecic},
  {Britt}, {Bowman}, {Wheatley}, {Loredo}, {Deming}, {Hebb}, {Hellier},
  {Maxted}, {Pollaco}, \& {West}}]{2010arXiv1003.2763C}
{Campo}, C.~J., {et~al.} 2010, ArXiv e-prints

\bibitem[{{Carter} \& {Winn}(2009)}]{2009ApJ...704...51C}
{Carter}, J.~A., \& {Winn}, J.~N. 2009, \apj, 704, 51

\bibitem[{{Carter} {et~al.}(2009){Carter}, {Winn}, {Gilliland}, \&
  {Holman}}]{2009ApJ...696..241C}
{Carter}, J.~A., {Winn}, J.~N., {Gilliland}, R., \& {Holman}, M.~J. 2009, \apj,
  696, 241

\bibitem[{{Charbonneau} {et~al.}(2000){Charbonneau}, {Brown}, {Latham}, \&
  {Mayor}}]{2000ApJ...529L..45C}
{Charbonneau}, D., {Brown}, T.~M., {Latham}, D.~W., \& {Mayor}, M. 2000, \apjl,
  529, L45

\bibitem[{{Claret}(2004)}]{2004yCat..34281001C}
{Claret}, A. 2004, VizieR Online Data Catalog, 342, 81001

\bibitem[{{Droege} {et~al.}(2006){Droege}, {Richmond}, {Sallman}, \&
  {Creager}}]{2006PASP..118.1666D}
{Droege}, T.~F., {Richmond}, M.~W., {Sallman}, M.~P., \& {Creager}, R.~P. 2006,
  \pasp, 118, 1666

\bibitem[{{Fortney} {et~al.}(2007){Fortney}, {Marley}, \&
  {Barnes}}]{2007ApJ...659.1661F}
{Fortney}, J.~J., {Marley}, M.~S., \& {Barnes}, J.~W. 2007, \apj, 659, 1661

\bibitem[{{Fortney} \& {Nettelmann}(2010)}]{2010SSRv..152..423F}
{Fortney}, J.~J., \& {Nettelmann}, N. 2010, \ssr, 152, 423

\bibitem[{Gelman \& Rubin(1992)}]{GR1992}
Gelman, A., \& Rubin, D.~B. 1992, Statistical Science, 7, 457

\bibitem[{{Gibson} {et~al.}(2010){Gibson}, {Pollacco}, {Barros}, {Benn},
  {Christian}, {Hrudkov{\'a}}, {Joshi}, {Keenan}, {Simpson}, {Skillen},
  {Steele}, \& {Todd}}]{2010MNRAS.401.1917G}
{Gibson}, N.~P., {et~al.} 2010, \mnras, 401, 1917

\bibitem[{{Hebb} {et~al.}(2009){Hebb}, {Collier-Cameron}, {Loeillet},
  {Pollacco}, {H{\'e}brard}, {Street}, {Bouchy}, {Stempels}, {Moutou},
  {Simpson}, {Udry}, {Joshi}, {West}, {Skillen}, {Wilson}, {McDonald},
  {Gibson}, {Aigrain}, {Anderson}, {Benn}, {Christian}, {Enoch}, {Haswell},
  {Hellier}, {Horne}, {Irwin}, {Lister}, {Maxted}, {Mayor}, {Norton}, {Parley},
  {Pont}, {Queloz}, {Smalley}, \& {Wheatley}}]{2009ApJ...693.1920H}
{Hebb}, L., {et~al.} 2009, \apj, 693, 1920

\bibitem[{{Henry} {et~al.}(2000){Henry}, {Marcy}, {Butler}, \&
  {Vogt}}]{2000ApJ...529L..41H}
{Henry}, G.~W., {Marcy}, G.~W., {Butler}, R.~P., \& {Vogt}, S.~S. 2000, \apjl,
  529, L41

\bibitem[{{Holman} {et~al.}(2006){Holman}, {Winn}, {Latham}, {O'Donovan},
  {Charbonneau}, {Bakos}, {Esquerdo}, {Hergenrother}, {Everett}, \&
  {P{\'a}l}}]{2006ApJ...652.1715H}
{Holman}, M.~J., {et~al.} 2006, \apj, 652, 1715

\bibitem[{{Madhusudhan} {et~al.}(2011){Madhusudhan}, {Harrington}, {Stevenson},
  {Nymeyer}, {Campo}, {Wheatley}, {Deming}, {Blecic}, {Hardy}, {Lust},
  {Anderson}, {Collier-Cameron}, {Britt}, {Bowman}, {Hebb}, {Hellier},
  {Maxted}, {Pollacco}, \& {West}}]{2011Natur.469...64M}
{Madhusudhan}, N., {et~al.} 2011, \nat, 469, 64

\bibitem[{{Mandel} \& {Agol}(2002)}]{2002ApJ...580L.171M}
{Mandel}, K., \& {Agol}, E. 2002, \apjl, 580, L171

\bibitem[{{Mandushev} {et~al.}(2007){Mandushev}, {O'Donovan}, {Charbonneau},
  {Torres}, {Latham}, {Bakos}, {Dunham}, {Sozzetti}, {Fern{\'a}ndez},
  {Esquerdo}, {Everett}, {Brown}, {Rabus}, {Belmonte}, \&
  {Hillenbrand}}]{2007ApJ...667L.195M}
{Mandushev}, G., {et~al.} 2007, \apjl, 667, L195

\bibitem[{{P{\'a}l}(2008)}]{2008MNRAS.390..281P}
{P{\'a}l}, A. 2008, \mnras, 390, 281

\bibitem[{{Perna} {et~al.}(2010){Perna}, {Menou}, \&
  {Rauscher}}]{2010ApJ...724..313P}
{Perna}, R., {Menou}, K., \& {Rauscher}, E. 2010, \apj, 724, 313

\bibitem[{{Pont} {et~al.}(2006){Pont}, {Zucker}, \&
  {Queloz}}]{2006MNRAS.373..231P}
{Pont}, F., {Zucker}, S., \& {Queloz}, D. 2006, \mnras, 373, 231

\bibitem[{{Sozzetti} {et~al.}(2007){Sozzetti}, {Torres}, {Charbonneau},
  {Latham}, {Holman}, {Winn}, {Laird}, \& {O'Donovan}}]{2007ApJ...664.1190S}
{Sozzetti}, A., {Torres}, G., {Charbonneau}, D., {Latham}, D.~W., {Holman},
  M.~J., {Winn}, J.~N., {Laird}, J.~B., \& {O'Donovan}, F.~T. 2007, \apj, 664,
  1190

\bibitem[{{Sozzetti} {et~al.}(2009){Sozzetti}, {Torres}, {Charbonneau}, {Winn},
  {Korzennik}, {Holman}, {Latham}, {Laird}, {Fernandez}, {O'Donovan},
  {Mandushev}, {Dunham}, {Everett}, {Esquerdo}, {Rabus}, {Belmonte}, {Deeg},
  {Brown}, {Hidas}, \& {Baliber}}]{2009ApJ...691.1145S}
{Sozzetti}, A., {et~al.} 2009, \apj, 691, 1145

\bibitem[{{Torres} {et~al.}(2008){Torres}, {Winn}, \&
  {Holman}}]{2008ApJ...677.1324T}
{Torres}, G., {Winn}, J.~N., \& {Holman}, M.~J. 2008, \apj, 677, 1324

\bibitem[{{Torres} {et~al.}(2007){Torres}, {Bakos}, {Kov{\'a}cs}, {Latham},
  {Fern{\'a}ndez}, {Noyes}, {Esquerdo}, {Sozzetti}, {Fischer}, {Butler},
  {Marcy}, {Stefanik}, {Sasselov}, {L{\'a}z{\'a}r}, {Papp}, \&
  {S{\'a}ri}}]{2007ApJ...666L.121T}
{Torres}, G., {et~al.} 2007, \apjl, 666, L121

\bibitem[{{Winn}(2010)}]{2010arXiv1001.2010W}
{Winn}, J.~N. 2010, ArXiv e-prints

\bibitem[{{Winn} {et~al.}(2007{\natexlab{a}}){Winn}, {Holman}, \&
  {Fuentes}}]{2007AJ....133...11W}
{Winn}, J.~N., {Holman}, M.~J., \& {Fuentes}, C.~I. 2007{\natexlab{a}}, \aj,
  133, 11

\bibitem[{{Winn} {et~al.}(2007{\natexlab{b}}){Winn}, {Holman}, {Bakos},
  {P{\'a}l}, {Johnson}, {Williams}, {Shporer}, {Mazeh}, {Fernandez}, {Latham},
  \& {Gillon}}]{2007AJ....134.1707W}
{Winn}, J.~N., {et~al.} 2007{\natexlab{b}}, \aj, 134, 1707

\bibitem[{{Winn} {et~al.}(2008){Winn}, {Holman}, {Torres}, {McCullough},
  {Johns-Krull}, {Latham}, {Shporer}, {Mazeh}, {Garcia-Melendo}, {Foote},
  {Esquerdo}, \& {Everett}}]{2008ApJ...683.1076W}
---. 2008, \apj, 683, 1076

\bibitem[{{Yi} {et~al.}(2001){Yi}, {Demarque}, {Kim}, {Lee}, {Ree}, {Lejeune},
  \& {Barnes}}]{2001ApJS..136..417Y}
{Yi}, S., {Demarque}, P., {Kim}, Y., {Lee}, Y., {Ree}, C.~H., {Lejeune}, T., \&
  {Barnes}, S. 2001, \apjs, 136, 417

\end{thebibliography}

\tabletypesize{\scriptsize}

\begin{deluxetable*}{lcccccl}
\tablecaption{Journal of Observations\label{tbl:obs}}
\tablewidth{0pt}

\tablehead{
\colhead{Date}&
\colhead{Target}&
\colhead{Filter}&
\colhead{Cadence}&
\colhead{RMS}&
\colhead{Red noise factor}&
\colhead{Notes} \\
\colhead{[UT]}&
\colhead{}&
\colhead{}&
\colhead{$\Gamma~[\rm min^{-1}]$}&
\colhead{$\sigma$}&
\colhead{$\beta$}&
\colhead{}
}

\startdata
08 Mar 2008 & HAT-P-3 & $i$ & 1.54 & 0.0018 & 1.39 &  \\
06 Apr 2008 & HAT-P-3 & $i$ & 2.50 & 0.0019 & 1.48 &  \\
05 May 2008 & HAT-P-3 & $i$ & 2.50 & 0.0036 & 1.00 & Incomplete phase coverage\\
18 Jan 2009 & HAT-P-3 & $z$ & 0.58 & 0.55   & --   & Bad weather\tablenotemark{a}\\
14 Mar 2009 & HAT-P-3 & $z$ & 0.81 & 0.0010 & 1.31 &  \\
20 Mar 2009 & HAT-P-3 & $z$ & 0.67 & 0.0014 & 1.24 & Incomplete phase coverage\\
15 Apr 2009 & HAT-P-3 & $z$ & 0.81 & 0.0012 & 1.05 &  \\
18 Apr 2009 & HAT-P-3 & $z$ & 0.81 & 0.0013 & --   & Bad weather\tablenotemark{a}\\
30 Jan 2010 & HAT-P-3 & $z$ & 0.58 & 0.0029 & --   & Very incomplete phase coverage\tablenotemark{a}\\
10 Jun 2008 & TrES-4  & $i$ & 0.81 & 0.0013 & 1.06 & Incomplete phase coverage\\
01 Apr 2009 & TrES-4  & $i$ & 0.81 & 0.0012 & 1.00 & \\
03 May 2009 & TrES-4  & $i$ & 0.67 & 0.0009 & 1.42 & Incomplete phase coverage\\
28 May 2009 & TrES-4  & $i$ & 0.71 & 0.0011 & --   & Very incomplete phase coverage\tablenotemark{a}\\
06 Jul 2009 & TrES-4  & $i$ & 0.67 & 0.0013 & --   & Very incomplete phase coverage\tablenotemark{a}\\
27 Apr 2010 & TrES-4  & $i$ & 0.45 & 0.0016 & 1.00 & \\
04 May 2010 & TrES-4  & $i$ & 0.67 & 0.0012 & 1.15 & Partial transit only\\
16 Jun 2010 & TrES-4  & $i$ & 0.81 & 0.0011 & --   & Very incomplete phase coverage\tablenotemark{a}\\
18 Dec 2008 & WASP-12 & $z$ & 0.80 & 0.0020 & --   & Bad weather\tablenotemark{a}\\
08 Jan 2009 & WASP-12 & $z$ & 1.50 & 0.0018 & 1.48 & \\
18 Jan 2009 & WASP-12 & $z$ & 0.67 & 0.0019 & --   & Bad weather\tablenotemark{a}\\
19 Jan 2009 & WASP-12 & $z$ & 1.00 & 0.0011 & --   & Bad weather\tablenotemark{a}\\
07 Mar 2009 & WASP-12 & $z$ & 0.80 & 0.0012 & --   & Bad weather\tablenotemark{a}\\
06 Dec 2009 & WASP-12 & $V$ & 6.33 & 0.0011 & 1.57 & Observed using the NOT\\
12 Jan 2010 & WASP-12 & $z$ & 1.00 & 0.0020 & --   & Bad weather\tablenotemark{a}\\
24 Jan 2010 & WASP-12 & $z$ & 0.50 & 0.0031 & --   & Bad weather\tablenotemark{a}\\
25 Jan 2010 & WASP-12 & $z$ & 0.50 & 0.0023 & --   & Bad weather\tablenotemark{a}\\
18 Feb 2010 & WASP-12 & $z$ & 0.40 & 0.0018 & --   & Bad weather\tablenotemark{a}\\
01 Mar 2010 & WASP-12 & $z$ & 0.80 & 0.0049 & --   & Bad weather\tablenotemark{a}\\
\enddata
\tablenotetext{a}{Data not used in calculations.}

\end{deluxetable*}

\begin{deluxetable*}{cccc}
\tablecaption{Photometry of TrES-4 (Excerpt)\label{tbl:photometry_tres4}}
\tablewidth{0pt}

\tablehead{
\colhead{Filter} &
\colhead{HJD$_{\rm UTC}$} & 
\colhead{Relative flux} &
\colhead{Airmass}
}

\startdata
i&	2454628.84123771&	1.00133657&	1.0055\\
i&	2454628.84212893&	0.99907639&	1.0058\\
i&	2454628.84384188&	0.99966428&	1.0062
\enddata 

\tablecomments{The time-stamp represents the UT-based Heliocentric Julian Date at midexposure.
We intend for the rest of this table to be available online.}

\end{deluxetable*}

\begin{deluxetable*}{cccc}
%\tabletypesize{\normalsize}
\tabletypesize{\scriptsize}
\tablecaption{Photometry of HAT-P-3 (Excerpt)\label{tbl:photometry_hat3}}
\tablewidth{0pt}

\tablehead{
\colhead{Filter} &
\colhead{HJD$_{\rm UTC}$} & 
\colhead{Relative flux} &
\colhead{Airmass}
}

\startdata
i&	2454534.74665524&	1.00252886&	1.5209\\
i&	2454534.74709502&	0.99836326&	1.5175\\
i&	2454534.74755801&	1.00316729&	1.5142
\enddata 

\tablecomments{The time-stamp represents the UT-based Heliocentric Julian Date at midexposure.
We intend for the rest of this table to be available online.}

\end{deluxetable*}

\begin{deluxetable*}{cccc}
%\tabletypesize{\normalsize}
\tabletypesize{\scriptsize}
\tablecaption{Photometry of WASP-12 (Excerpt)\label{tbl:photometry_wasp12}}
\tablewidth{0pt}

\tablehead{
\colhead{Filter} &
\colhead{HJD$_{\rm UTC}$} & 
\colhead{Relative flux} &
\colhead{Airmass}
}

\startdata
z&	2454840.62025635&	1.00042427&	1.5419\\
z&	2454840.62085820&	1.00001333&	1.5362\\
z&	2454840.62148318&	1.00215610&	1.5304
\enddata 

\tablecomments{The time-stamp represents the UT-based Heliocentric Julian Date at midexposure.
We intend for the rest of this table to be available online.}

\end{deluxetable*}

\tabletypesize{\normalsize}

\begin{deluxetable}{lcccl}
\tablecaption{System Parameters for TrES-4\label{tbl:params-tres4}}
\tablewidth{0pt}

\tablehead{
\colhead{Parameter} & \colhead{Symbol} & \colhead{Value} & \colhead{68.3\% Conf.~Limits} & \colhead{Comment\tablenotemark{a}}
}

\startdata
{\it Transit parameters:} & & & \\
Orbital period~[d]&			$P$&		$3.5539268$&		$\pm 0.0000032$&	LC\\
Midtransit time~[HJD$_{\rm UTC}$]&			$T_0$&		$2454230.90524$&	$\pm 0.00062$&		LC\\
Planet-to-star radius ratio&		$R_p/R_\star$&	$0.09745$&		$\pm 0.00076$&		LC\\
Orbital inclination~[deg]&		$i$&		$82.81$&		$\pm 0.37$&		LC\\
Scaled semimajor axis&			$a/R_\star$&	$6.08$&			$\pm 0.16$&		LC\\
Transit impact parameter&		$b$&		$0.761$&		$\pm 0.018$&		LC\\
Transit duration~[hr]&			&		$3.567$&		$\pm 0.037$&		LC\\
Transit ingress or egress duration~[hr]&&		$0.692$&		$\pm 0.047$&		LC\\
& & & \\
{\it Other orbital parameters:} & & & \\
Orbital eccentricity&			$e$&		$0$&			&			Adopted\\
Velocity semiamplitude~[m~s$^{-1}$]&	$K$&		$97.4$&			$\pm 7.2$&		K\\
Planet-to-star mass ratio&		$M_p/M_\star$&	$0.000631$&		$\pm 0.000047$&		LC + M + T + Y$^2$ + K\\
Semimajor axis~[AU]&			$a$&		$0.05084$&		$\pm 0.00050$&		LC + M + T + Y$^2$\\
& & & \\
{\it Stellar parameters:} & & & \\
Mass~[M$_{\odot}$]& $M_\star$&				$1.388$&		$\pm 0.042$&		LC + M + T + Y$^2$\\
Radius~[R$_{\odot}$]& $R_\star$&			$1.798$&		$\pm 0.052$&		LC + M + T + Y$^2$\\
Mean density~[g~cm$^{-3}$]& $\rho_\star$&		$0.337$&		$\pm 0.027$&		LC\\
Effective temperature~[K]& $T_{\rm eff}$&		$6200$&			$\pm 75$&		T\\
Projected rotation rate~[km~s$^{-1}$]&	$v\sin i_\star$&$9.5$&			$\pm 1.0$&		PRR\\
Surface gravity~[cgs]&			$\log g_\star$&	$4.071$&		$\pm 0.024$&		LC + M + T + Y$^2$\\
Metallicity~[dex]&			[Fe/H]&		$+0.14$&		$\pm 0.09$&		M\\
Luminosity~[L$_{\odot}$]&		$L_\star$&	$4.25$&			$\pm 0.41$&		LC + M + T + Y$^2$\\
Visual Magnitude~[mag]&			M$_{\rm V}$&	$3.19$&			$\pm 0.12$&		LC + M + T + Y$^2$\\
Apparent Magnitude~[mag]&		m$_{\rm V}$&	$11.592$&		$\pm 0.004$&		m$_{\rm V}$\\
Age~[Gyr]&				&		$2.9$&			$\pm 0.3$&		LC + M + T + Y$^2$\\
Distance~[pc]&				&		$479$&			$\pm 26$&		LC + M + T + Y$^2$ + m$_{\rm V}$\\
& & & \\
{\it Planetary parameters:} & & & \\
Mass~[M$_{\rm Jup}$]&			$M_p$&		$0.917$&		$\pm 0.070$&		LC + M + T + Y$^2$ + K\\
Radius~[R$_{\rm Jup}$]&			$R_ p $&	$1.706$&		$\pm 0.056$&		LC + M + T + Y$^2$\\
Surface gravity~[cgs]&			$\log g_p$&	$2.893$&		$\pm 0.042$&		LC + K\\
Mean density~[g~cm$^{-3}$]&		$\rho_p$&	$0.229$&		$\pm 0.027$&		LC + M + T + Y$^2$
\enddata

\tablenotetext{a}{The type of data used to calculate the given quantity.  LC denotes light curve data. M and T denote metallicity and temperature respectively and were taken from \citet{2009ApJ...691.1145S}. K, PRR, and m$_{\rm V}$ denote velocity semiamplitude, projected rotation rate, and apparent magnitude respectively and were taken from \citet{2007ApJ...667L.195M}.  Y$^2$ denotes the Yonsei-Yale isochrones \citep{2001ApJS..136..417Y}}
\end{deluxetable}

\tabletypesize{\normalsize}

\begin{deluxetable}{lcccl}
\tablecaption{System Parameters for HAT-P-3\label{tbl:params-hat3}}
\tablewidth{0pt}

\tablehead{
\colhead{Parameter} & \colhead{Symbol} & \colhead{Value} & \colhead{68.3\% Conf.~Limits} & \colhead{Comment\tablenotemark{a}}
}

\startdata
{\it Transit parameters:} & & & \\
Orbital period~[d]&			$P$&		$2.8997360$&		$\pm 0.0000020$&	LC\\
Midtransit time~[HJD$_{\rm UTC}$]&			$T_0$&		$2454856.70118$&	$\pm 0.00010$&		LC\\
Planet-to-star radius ratio&		$R_p/R_\star$&	$0.1063$&		$\pm 0.0020$&		LC\\
Orbital inclination~[deg]&		$i$&		$87.07$&		$\pm 0.55$&		LC\\
Scaled semimajor axis&			$a/R_\star$&	$10.39$&		$\pm 0.49$&		LC\\
Transit impact parameter&		$b$&		$0.530$&		$\pm 0.075$&		LC\\
Transit duration~[hr]&			&		$2.075$&		$\pm 0.022$&		LC\\
Transit ingress or egress duration~[hr]&&		$0.270$&		$\pm 0.033$&		LC\\
& & &\\
{\it Other orbital parameters:} & & &\\
Orbital eccentricity&			$e$&		$0$&			&			Adopted\\
Velocity semiamplitude~[m~s$^{-1}$]&	$K$&		$89.1$&			$\pm 2.0$&		K\\
Planet-to-star mass ratio&		$M_p/M_\star$&	$0.000615$&		$\pm 0.000015$&		LC + M + T + Y$^2$ + K\\
Semimajor axis~[AU]&			$a$&		$0.03866$&		$\pm 0.00041$&		LC + M + T + Y$^2$\\
& & &\\
{\it Stellar parameters:} & & &\\
Mass~[M$_{\odot}$]&			$M_\star$&	$0.917$&		$\pm 0.030$&		LC + M + T + Y$^2$\\
Radius~[R$_{\odot}$]&			$R_\star$&	$0.799$&		$\pm 0.039$&		LC + M + T + Y$^2$\\
Mean density~[g~cm$^{-3}$]&		$\rho_\star$&	$2.53$&			$\pm 0.36$&		LC\\
Effective temperature~[K]&		$T_{\rm eff}$&	$5185$&			$\pm 80$&		T\\
Projected rotation rate~[km~s$^{-1}$]&	$v\sin i_\star$&$0.5$&			$\pm 0.5$&		PRR\\
Surface gravity~[cgs]&			$\log g_\star$&	$4.594$&		$\pm 0.041$&		LC + M + T + Y$^2$\\
Metallicity~[dex]&			[Fe/H]&		$+0.27$&		$\pm 0.08$&		M\\
Luminosity~[L$_{\odot}$]&		$L_\star$&	$0.435$&		$\pm 0.053$&		LC + M + T + Y$^2$\\
Visual Magnitude~[mag]&			M$_{\rm V}$&	$5.87$&			$\pm 0.15$&		LC + M + T + Y$^2$\\
Apparent Magnitude~[mag]&		m$_{\rm V}$&	$11.577$&		$\pm 0.067$&		m$_{\rm V}$\\
Age [Gyr]&				&		$1.6$&			$-1.3, +2.9$&		LC + M + T + Y$^2$\\
Distance [pc]&				&		$138$&			$\pm 10$&		LC + M + T + Y$^2$ + m$_{\rm V}$\\
&&&\\
{\it Planetary parameters:}&&&\\
Mass~[M$_{\rm Jup}$]&			$M_p$&		$0.591$&		$\pm 0.018$&		LC + M + T + Y$^2$ + K\\
Radius~[R$_{\rm Jup}$]&			$R_p$&		$0.827$&		$\pm 0.055$&		LC + M + T + Y$^2$\\
Surface gravity~[cgs]&			$\log g_p$&	$3.330$&		$\pm 0.058$&		LC + K\\
Mean density~[g~cm$^{-3}$]&		$\rho_p$&	$1.29$&			$\pm 0.25$&		LC + M + T + Y$^2$
\enddata

\tablenotetext{a}{The type of data used to calculate the given quantity.  LC denotes light curve data, K, M, T, and PRR denote velocity semiamplitude, metallicity, temperature, and projected rotation rate respectively and were taken from \citet{2007ApJ...666L.121T}.  The uncertanties in M and T were adjusted as suggested by \citet{2008ApJ...677.1324T}.  m$_{\rm V}$ denotes apparent magnitude and was taken from \citet{2006PASP..118.1666D}.  Y$^2$ denotes the Yonsei-Yale isochrones \citep{2001ApJS..136..417Y}.}
\end{deluxetable}

\tabletypesize{\normalsize}

\begin{deluxetable}{lcccl}
\tablecaption{System Parameters for WASP-12\label{tbl:params-wasp12}}
\tablewidth{0pt}

\tablehead{
\colhead{Parameter} & \colhead{Symbol} & \colhead{Value} & \colhead{68.3\% Conf.~Limits} & \colhead{Comment\tablenotemark{a}}
}

\startdata
{\it Transit parameters:}&&&\\
Orbital period~[d]&		$P$&			$1.0914222$&		$\pm 0.0000011$&	LC\\
Midtransit time~[HJD$_{\rm UTC}$]&		$T_0$&			$2454508.97605$&	$\pm 0.00028$&		LC\\
Planet-to-star radius ratio&	$R_p/R_\star$&		$0.1119$&		$\pm 0.0020$&		LC\\
Orbital inclination~[deg]&	$i$&			$86.0$&			$\pm 3.0$&		LC\\
Scaled semimajor axis&		$a/R_\star$&		$3.097$&		$\pm 0.082$&		LC\\
Transit impact parameter&	$b$&			$0.22$&			$\pm 0.15$&		LC\\
Transit duration~[hr]&		&			$3.001$&		$\pm 0.037$&		LC\\
Transit ingress or egress duration~[hr]&&		$0.324$&		$\pm 0.020$&		LC\\
&&&\\
{\it Other orbital parameters:}&&&\\
Orbital eccentricity&		$e$&			$0$&			&			Adopted\\
Velocity semiamplitude~[m~s$^{-1}$]&	$K$&		$226$&			$\pm 4$&		K\\
Planet-to-star mass ratio&	$M_p/M_\star$&		$0.000993$&		$\pm 0.000038$&		LC + M + T + Y$^2$ + K\\
Semimajor axis~[AU]&		$a$&			$0.02293$&		$\pm 0.00078$&		LC + M + T + Y$^2$\\
&&&\\
{\it Stellar parameters:}&&&\\
Mass~[M$_{\odot}$]&		$M_\star$&		$1.35$&			$\pm 0.14$&		LC + M + T + Y$^2$\\
Radius~[R$_{\odot}$]&		$R_\star$&		$1.599$&		$\pm 0.071$&		LC + M + T + Y$^2$\\
Mean density~[g~cm$^{-3}$]&	$\rho_\star$&		$0.472$&		$\pm 0.038$&		LC\\
Effective temperature~[K]&	$T_{\rm eff}$&		$6300$&			$\pm 150$&		T\\
Projected rotation rate~[km~s$^{-1}$]&	$v\sin i_\star$&$<2.2$&			$\pm 1.5$&		PRR\\
Surface gravity~[cgs]&		$\log g_\star$&		$4.162$&		$\pm 0.029$&		LC + M + T + Y$^2$\\
Metallicity~[dex]&		[Fe/H]&			$0.30$&			$\pm 0.10$&		M\\
Luminosity~[L$_{\odot}$]&	$L_\star$&		$3.0$&			$\pm 1.2$&		LC + M + T + Y$^2$\\
Visual Magnitude~[mag]&		M$_{\rm V}$&		$3.54$&			$\pm 0.38$&		LC + M + T + Y$^2$\\
Apparent Magnitude~[mag]&	m$_{\rm V}$&		$11.69$&		$\pm 0.08$&		m$_{\rm V}$\\
Age [Gyr]&			&			$1.7$&			$\pm 0.8$&		LC + M + T + Y$^2$\\
Distance [pc]&			&			$427$&			$\pm 90$&		LC + M + T + Y$^2$ + m$_{\rm V}$\\
&&&\\
{\it Planetary parameters:}&&&\\
Mass~[M$_{\rm Jup}$]&		$M_p$&			$1.404$&		$\pm 0.099$&		LC + M + T + Y$^2$ + K\\
Radius~[R$_{\rm Jup}$]&		$R_p$&			$1.736$&		$\pm 0.092$&		LC + M + T + Y$^2$\\
Surface gravity~[cgs]&		$\log g_p$&		$3.066$&		$\pm 0.031$&		LC + K\\
Mean density~[g~cm$^{-3}$]&	$\rho_p$&		$0.337$&		$\pm 0.039$&		LC + M + T + Y$^2$
\enddata

\tablenotetext{a}{The type of data used to calculate the given quantity.  LC denotes light curve data. K, M, T, PRR, and m$_{\rm V}$ denote velocity semiamplitude, metallicity, temperature, projected rotation rate and apparent magnitude respectively, and were taken from \citet{2009ApJ...693.1920H} (the uncertainties were symmetrized). Y$^2$ denotes the Yonsei-Yale isochrones \citep{2001ApJS..136..417Y}.}
\end{deluxetable}

\begin{deluxetable*}{ccccl}
\tablecaption{Midtransit Times for TrES-4\label{tbl:midtransits-tres4}}
\tablewidth{0pt}

\tablehead{
\colhead{Epoch}&\colhead{Midtransit Time [HJD]}  &
\colhead{Uncertainty}   & \colhead{Filter} & \colhead{Source}
}

\startdata
  -2  &		2454223.797215 & 0.000847 & z & \citet{2009ApJ...691.1145S}\tablenotemark{a}\\
   0  &		2454230.904913 & 0.000656 & z & \citet{2009ApJ...691.1145S}\tablenotemark{a}\\
   0  &		2454230.905624 & 0.001106 & B & \citet{2009ApJ...691.1145S}\tablenotemark{a}\\
 112  &		2454628.947249 & 0.001056 & i & This paper\tablenotemark{b}\\ %2008 Jun 10
 195  &		2454923.920285 & 0.000492 & i & This paper\\ %2009 Apr 01
 204  &		2454955.907886 & 0.001138 & i & This paper\tablenotemark{b}\\ %2009 May 03
 305  &		2455314.853547 & 0.000616 & i & This paper\\ %2010 Apr 27
 307  &		2455321.959132 & 0.001049 & i & This paper\tablenotemark{b} % 2010 May 04
\enddata

\tablenotetext{a}{Transits were observed by \citet{2007ApJ...667L.195M} but re-analyzed by \cite{2009ApJ...691.1145S}}
\tablenotetext{b}{Partial transit observed}
\end{deluxetable*}

\begin{deluxetable*}{ccccl}
\tablecaption{Midtransit Times for HAT-P-3\label{tbl:midtransits-hat3}}
\tablewidth{0pt}

\tablehead{
\colhead{Epoch} & \colhead{Midtransit Time [HJD]}  &
\colhead{Uncertainty}   & \colhead{Filter} & \colhead{Source}
}

\startdata
-220  &  2454218.759400 & 0.002900 &i&  \citet{2007ApJ...666L.121T}\\
-111  &  2454534.830232 & 0.000391 &i&  This paper\\ %2008 Mar 08\\
-101  &  2454563.828281 & 0.000358 &i&  This paper\\ %2008 Apr 06\\
 -91  &  2454592.824589 & 0.000707 &i&  This paper\tablenotemark{a}\\ %2008 May 05\\
   0  &  2454856.701370 & 0.000240 &--\tablenotemark{b}&  \citet{2010MNRAS.401.1917G}\\
   1  &  2454859.600240 & 0.000370 &--\tablenotemark{b}&  \citet{2010MNRAS.401.1917G}\\
  17  &  2454905.996437 & 0.000385 &z& This paper\\ %2009 Mar 14\\
  19  &  2454911.795918 & 0.000510 &z& This paper\tablenotemark{a}\\ %2009 Mar 20\\
  22  &  2454920.495670 & 0.000550 &--\tablenotemark{b}&  \citet{2010MNRAS.401.1917G}\\
  28  &  2454937.893861 & 0.000317 &z& This paper\\ %2009 Apr 15\\
  31  &  2454946.592600 & 0.000650 &--\tablenotemark{b}&  \citet{2010MNRAS.401.1917G}\\
  32  &  2454949.493340 & 0.000400 &--\tablenotemark{b}&  \citet{2010MNRAS.401.1917G}\\
  41  &  2454975.590370 & 0.000340 &--\tablenotemark{b}&  \citet{2010MNRAS.401.1917G}\\
  42  &  2454978.489930 & 0.000510 &--\tablenotemark{b}&  \citet{2010MNRAS.401.1917G}
\enddata

\tablenotetext{a}{Partial transit observed}
\tablenotetext{b}{A single wideband filter ($\approx 500-700$nm) was used.}
\end{deluxetable*}

\begin{deluxetable*}{ccccl}
\tablecaption{Midtransit Times for WASP-12\label{tbl:midtransits-wasp12}}
\tablewidth{0pt}

\tablehead{
\colhead{Epoch} & \colhead{Midtransit Time [HJD]}  &
\colhead{Uncertainty}   & \colhead{Filter} & \colhead{Source}
}

\startdata
   0  &  2454508.97610 & 0.00020 & $\cdots$ &  \citet{2009ApJ...693.1920H}\tablenotemark{a}\\
 304  &  2454840.76781 & 0.00047 &V&  This paper\\
 608  &  2455172.56103 & 0.00044 &V&  This paper
\enddata

\tablenotetext{a}{This data point actually represents a consensus value derived from
observations of several transits.}
\end{deluxetable*}

\end{document}